# RUNNING COUPLING CONSTANTS IN 2D GRAVITY

Christof Schmidhuber

*California Institute of Technology, Pasadeana, CA 91125*

The topic of this talk is the renormalization group flow in two–dimensional field theories that are coupled to gravity. I will explain the basic ideas at the example of the sine-Gordon model. I will define in a minute, what is meant by scale–dependent coupling constants in theories with gravity, where the scale itself is a dynamical variable. Before, it is necessary to generalize the theory of David, Distler and Kawai in order to describe two–dimensional quantum gravity coupled to *non*–conformal matter.

## 1. New Terms in the DDK Action

In David, Distler and Kawai's[1,2] approach, the action for 2D gravity on genus 0, coupled to a scalar matter field $x$ with central charge $c = 1$, is usually assumed as

$$S = \frac{1}{8\pi} \int \sqrt{\hat{g}} \{ (\partial x)^2 + (\partial \phi)^2 + 2\sqrt{2}\hat{R}\phi + \tau^i \, \Phi_i(x) \, e^{\alpha_i \phi} \\ + \text{cosmological constant} \}. \tag{1}$$

Here, $\hat{g}$ is the (fictitious) background metric, $\phi$ is the Liouville mode and the $\tau^i$ are dimensionless coupling constants (powers of a length scale $a$ are omitted here and below). The $\Phi_i(x)$ are scaling operators of the matter theory, such as the sine–Gordon interaction $\cos px$. They are "gravitationally dressed" by exponentials $e^{\alpha_i \phi}$, where $\alpha_i$ is adjusted so that the dimensions of the dressed operators are two. Summation over $i$ is understood in (1). We will ignore all problems associated with the presence of the cosmological constant. It can be argued that they have no qualitative effects on what follows.[3] This will be confirmed below by agreement with matrix model results.

As it stands, theory (1) has a problem: it is not scale invariant for finite $\tau^i$. It should be scale invariant, because in quantum gravity scale invariance is part of the gauge invariance. In particular, all beta functions in the theory should be zero to all orders. But in (1) they aren't: generally, the beta functions are given by[4]

$$\beta^i \equiv \dot\tau^i = (\Delta^i_j - 2\delta^i_j) \, \tau^j + \pi c^i_{jk} \tau^j \tau^k + O(\tau^3). \tag{2}$$

Here, $\Delta^i_j$ is the dimension matrix computed with $S_0$, and $c^i_{jk}$ are the operator product coefficients. It is clear that by adjusting $\alpha_i$ in (1) we can make the linear terms in

(2) vanish: $\Delta^i_j = 2\delta^i_j$.[1,2] But whenever there are nontrivial $c^i_{jk}$'s, the beta functions have quadratic pieces. By "nontrivial $c^i_{jk}$'s" I mean universal ones, that cannot be eliminated by picking a particular renormalization scheme. In the $c = 1$ model, universal $c^i_{jk}$'s are present whenever the interactions in (1) are "discrete primary fields", like $\cos px$ near the "special momenta" $p = \sqrt{2}$ or $p = \sqrt{2}/2$.[5]

The only way to cure the problem of nonvanishing beta functions is to add new terms to the action (1), of the form[3]

$$-\pi\, c^k_{ij}\tau^i\tau^j \int X_k(x,\phi) \; + \; \text{higher orders}, \tag{3}$$

where the operators $X_k$ are, up to field redefinitions, uniquely determined by scale invariance at $O(\tau^2)$. Their form is simple, but let me just state the result for the example of the sine–Gordon model coupled to gravity, $\Phi_i(x) = \cos px$ with $p = \sqrt{2}+\epsilon$, $\epsilon$ being small. $p = \sqrt{2}$ is the momentum where the Kosterlitz–Thouless transition takes place in the sine–Gordon model without gravity. (1) plus (3) comes out to be:

$$\begin{aligned} S = \frac{1}{8\pi} \int \sqrt{\hat{g}}\{\partial x^2 + \partial\phi^2 + 2\sqrt{2}\hat{R}\phi + \text{cosmological constant}\} \\ + m \int \cos(\sqrt{2}+\epsilon)x\, e^{\epsilon\phi} - \frac{\pi}{8\sqrt{2}}\, m^2 \int \phi\, \partial x^2. \end{aligned} \tag{4}$$

## 2. Running Coupling Constants

I am now ready to come to my title – running coupling constants. We have just introduced new terms in order to insure that the coupling constants do *not* run – with respect to the background scale $\sqrt{\hat{g}}$. But the physical scale in DDK's approach is $\sqrt{\hat{g}}e^{\alpha\phi}$, $\alpha = -\sqrt{2}$. Therefore, a shift of $\phi$ by a constant $\lambda$,

$$\phi \to \phi + \lambda, \tag{5}$$

is a scale transformation (see refs. 3,6). So we can ask: how do we have to make $m$ and $\epsilon$ $\lambda$–dependent in (4), so as to absorb the shift (5). $m(\lambda)$ and $\epsilon(\lambda)$ are what I call "running coupling constants." They are easy to find in this example:

$$m(\lambda) = m_0 e^{-\epsilon\lambda}, \quad \epsilon(\lambda) = \epsilon_0 - \frac{\pi^2}{2}\lambda m^2. \tag{7}$$

Here, $m_0$ and $\epsilon_0$ are constants. In deriving $\epsilon(\lambda)$, the $\lambda m^2 \partial x^2$ term has been absorbed in a redefinition of $x$ and in a shift of $\epsilon$. Defining 'dot' as $\frac{d}{d\lambda}$, we get the lowest order "beta functions"

$$\dot{\epsilon} \sim -m^2, \quad \dot{m} \sim -\epsilon m. \tag{8}$$

The resulting flow diagram is shown in fig.1. It is qualitatively the same as the Kosterlitz–Thouless diagramm of the flat–space sine–Gordon model, although quantitatively it is modified by the effects of gravity. We see that the new term $\phi\,\partial x^2$ in (4) plays a crucial role: Via the shift of $\epsilon$ in (7), it causes the flow in the $\epsilon$ direction. Ignoring it would be like forgetting field renormalization in the ordinary sine–Gordon model. As a consequence of this term, there is a diagonal critical line at

$$\epsilon > 0, \quad |m| \sim \epsilon,$$

corresponding to a linear phase boundary. This fact agrees to this order in $m$ and $\epsilon$ with recent matrix model results by Moore,[7] who found a singularity of the free energy at

$$\epsilon > 0, \quad |m| \sim \epsilon\, e^{\frac{1}{2}\sqrt{2}\epsilon\log\epsilon}. \tag{9}$$

Thus we "see" the $O(m^2)$ term in the Liouville action (4) in the matrix model results. It will be interesting to see if the logarithm in (9) follows from further modifications of (4), needed to keep the interaction near $p = \sqrt{2}$ marginal beyond $O(m^2)$.

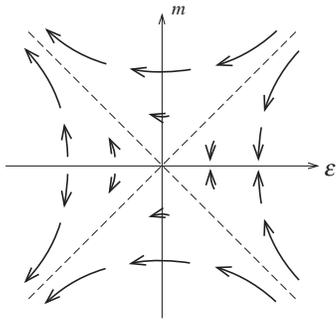
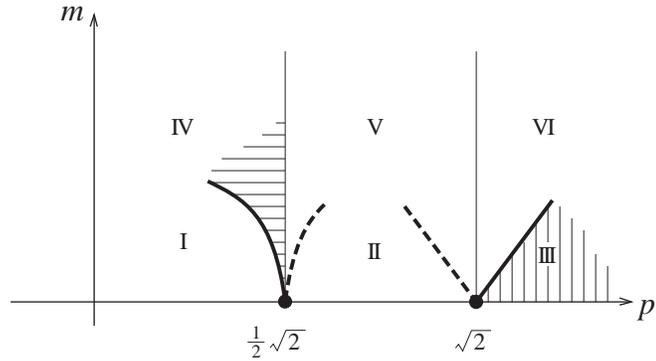

Fig.1: Kosterlitz–Thouless transition with gravity, to leading order; arrows point to the infrared.

Fig.2: Phase diagram for the sine–Gordon model coupled to gravity, to leading order.

## 3. Phase Diagram of the Sine–Gordon Model with Gravity

We can now interprete the phase diagram of ref. 7(fig. 2) near $m = 0, p = \sqrt{2}$: For $p < \sqrt{2}$ (regions II and V), $m$ grows exponentially in the IR. The $x$ field will thus get stuck in the minima of the sine–Gordon potential and the IR limit will be (infinitely many copies of) the $c = 0$, pure gravity model.[8] For $\epsilon > 0$ but $m$ greater than a critical value $m_c(\epsilon)$ (region VI), the IR limit is again the $c = 0$ model. For $m < m_c(\epsilon)$ (region III), the model flows to the free $c = 1$ model. However, the domain of small $\epsilon, m$ is now the IR domain, where the cosmological constant cannot be neglected and further investigation is needed.

As a second example of the correction terms (3), consider the sine–Gordon model in the vicinity of half the Kosterlitz–Thouless momentum, $p = \frac{1}{2}\sqrt{2} + \delta$. To quadratic order, the action is[3]

$$S = \frac{1}{8\pi} \int \sqrt{\hat{g}} \{\partial x^2 + \partial \phi^2 + 2\sqrt{2}\hat{R}\phi + \text{ghosts}\} \\ + m \int \cos(\frac{1}{2}\sqrt{2} + \delta)x \; e^{(-\frac{1}{2}\sqrt{2}+\delta)\phi} + (\frac{\mu}{8\pi} + \frac{m^2\pi}{8\delta}) \int \phi \; e^{-\sqrt{2}\phi}, \tag{10}$$

where we have used the form $\phi e^{-\sqrt{2}\phi}$ for the cosmological constant.[9] A cosmological constant term is induced at order $m^2$. It becomes comparable with the background cosmological constant at $\delta \sim m^2/\mu$. Although the situation is not entirely clear, we therefore expect some kind of phase transition at

$$\delta < 0, \quad |m| \sim \sqrt{|\mu\delta|},$$

where the last term in (10) becomes negative (fig.2). Indeed, in the matrix model a singularity of the free energy has been found at[7]

$$\delta < 0, \quad |m| \sim \sqrt{|\mu\delta|} \; e^{\frac{1}{2}\sqrt{2}\delta \log \delta}. \tag{12}$$

Let us therefore identify the region where $\mu + \frac{m^2}{\delta}\pi^2$ is negative with region IV of ref. 7. A further interpretation of the phase diagram must be left for the future work.

Let me summarize: There are new terms in the DDK action that describes two–dimensional quantum gravity coupled to nonconformal matter. These terms are required by scale invariance. They are crucial in deriving the renormalization group flow in the continuum theory. The results are consistent with the phase structure observed in the matrix model. Interestingly, we do not seem to make a big error by ignoring the problems associated with the presence of the cosmological constant.